# Correlation of structural, nanomechanical and electrostatic properties of single and few-layers MoS$_2$

Benjamin J. Robinson[1], Cristina E. Giusca[2], Yurema Teijeiro Gonzalez[2], Nicholas D. Kay[1], Olga Kazakova[2*] and Oleg V. Kolosov[1*]

[1]Department of Physics, Lancaster University, Lancaster, LA1 4YB, UK; email: o.kolosov@lancaster.ac.uk
[2]National Physical Laboratory, Hampton Road, Teddington, TW11 0LW, UK; email: olga.kazakova@npl.co.uk

**Abstract**

We have decoupled the intrinsic optical and electrostatic effects arising in monolayer and few-layer MoS$_2$ from those influenced by the flake-substrate interaction. Using ultrasonic force microscopy (UFM), we identify the change in the flake's nanomechanical properties for the suspended and supported regions on a trenched substrate. These regions are correlated with Raman spectra of the flake and its surface potential as measured by scanning Kelvin probe microscopy (SKPM). Relative to the supported region, we observe an increase in surface potential contrast due to suppressed charge transfer for the suspended monolayer. For the monolayer region, Raman spectra differ for the supported vs suspended areas. The study of the effect induced by different substrates reveals a red shift of the E$^1_{2g}$ mode for monolayer MoS$_2$ deposited on Si, consistent with a more strained MoS$_2$ on the Si substrate compared to the Au substrate.

**Introduction**

Layered transition metal dichalcogenides have attracted significant attention due to their potential applications in electronic and optical devices [1]. Molybdenum disulphide (MoS$_2$) is one of the most stable layered materials of this class. In the bulk form this semiconductor material has an indirect band gap of ~1.29 eV and is used in a broad range of diverse applications, *e.g.* as a photocatalyst and dry lubricant, as well as for photovoltaic power generation [2] and photo-electrochemical hydrogen and Li ion batteries production [3]. Monolayer MoS$_2$ has a ~1.75 eV direct band gap and prominent electro- and photoluminescent properties, making it a likely candidate for applications in photodetectors and light-emitting devices operating in the visible range [4]. Additionally, monolayer MoS$_2$-based field-effect transistors demonstrated very promising electronic characteristics, such as a large current on/off ratio and sub-threshold swing [5].

With a rapidly increasing interest in the development of ultrathin MoS$_2$-based devices, measurement methods allowing for multifunctional characterisation of physical properties, easy identification of MoS$_2$ layer number and interaction of the flakes with a substrate are in high demand. As electronic and optical properties of MoS$_2$ are strongly thickness and layer-substrate interaction dependent [6], it is essential to precisely ascribe the measured parameters to individual layers. Here, we have used Raman spectroscopy, scanning Kelvin probe microscopy (SKPM) and atomic and ultrasonic force microscopy (AFM/UFM) for the mapping of mechanically exfoliated MoS$_2$ flakes with domains of the different thickness with the aim to precisely correlate their optical, nanomechanical and electrostatic properties on the nanoscale as well as to explore the effect of interaction of MoS$_2$ flakes with a substrate.

Raman spectroscopy has been widely used to determine the number of graphene [7] and MoS$_2$ layers, strain in both materials [8, 9] and, in a bulk form, identification of various crystalline forms of the material [10, 11]. For off-resonant laser lines (*i.e.* 532 nm), there are four well-defined first-order Raman active modes, which can be typically observed in bulk MoS$_2$. The most prominent of them are the E$^1_{2g}$ mode (opposite in-plane vibration of two S atoms with respect to the Mo atom) and A$_{1g}$ mode (out-of-plane vibration of S atoms in opposite directions) [12]. It has been shown that for single- and few-layers MoS$_2$ both of these modes are thickness-dependent and shift away from each other in frequency with increasing thickness [13], hence providing a convenient and reliable means for determining layer thickness with atomic-level precision [14-17].

Similarly, the work function or surface potential of layered materials is also strongly dependent on the number of layers [18]. While the Raman studies of MoS$_2$ are relatively common, direct measurements of the screening length are comparatively rare and the role of the electrostatic coupling in layered materials is largely neglected, with an exception of graphene [19-21]. In previous works, thickness-dependent interlayer screening effects in MoS$_2$ were measured by means of Electrostatic Force Microscopy [22] and SKPM [23] techniques. For example, it has been demonstrated that the surface potential of pristine MoS$_2$ flakes decreased with increasing thickness (in tip-biased studies) and increasing with thickness (in sample-biased ones). Substrate related effects (*i.e.* trapped charges) and contamination caused by material processing can significantly affect the work function value [24, 25]. The estimated characteristic screening effects vary considerably in different studies, being within the length of ~5 nm (8 layers) [23] or ~30–50 nm [22]. In both cases the obtained screening length is significantly larger than in graphene, where a strong-coupling regime is achieved already at ~1–2 nm, albeit strongly depending on the initial charge density [18, 19]. However, in a recent experiment the screening length of ~2.96 nm was reported [24], making it similar to the one in graphene despite their significant differences in the conductivity and the electronic structure. Thus, the observed electrostatic properties of layered MoS$_2$ were attributed to a weak-coupling regime leading to the reduced screening properties.

UFM has been shown to be highly effective at determining the nanomechanical properties of MoS$_2$, arising from both intrinsic structure and defects as well as from the sample-substrate interface [4]. The inter-layer coupling in MoS$_2$ is largely dependent on layer stacking where properties such as folds can decrease the coupling strength by up to a factor of 5 [26]. This decrease in coupling strength between the layers will manifest itself as a decrease in the mechanical strength of the flake. The mechanical properties of MoS$_2$ can also be greatly affected by the substrate properties and morphology. It has been shown that MoS$_2$ deposited through mechanical exfoliation onto a layer of thermally oxidised SiO$_2$ does not follow its nanoscale rough surface but instead adheres to the high points on the substrate [4]. This surface roughness affects the coupling of the flake to the surface and as such may cause a local variance in the flake's mechanical and electrical properties [4, 25, 27]. Experimentally, variations in the Young's modulus and pre-tension in MoS$_2$ flakes have been observed and attributed to changing defect densities and adhesion to the substrate [27]. Using nanomechanical mapping by UFM [28, 29] one can observe the local mechanical properties, stresses [30] and adhesion of the multi-layer solid state structures [9, 31, 32] and, therefore probe the level of substrate-flake interaction and layer-layer interaction. Additionally, by studying suspended MoS$_2$ films it is possible to eliminate the substrate flake interaction altogether and observe purely the effect of interlayer coupling of the flake on its mechanical properties.

In this work we study the local optical, nanomechanical and electrostatic properties of single and few-layers MoS$_2$ as measured by a combination of functional scanning probe microscopy techniques and Raman spectroscopy mapping.

**Materials and Methods**

**Sample preparation**

The MoS$_2$ layers were produced by mechanical exfoliation from bulk crystals using the well-established 'scotch tape' method [33], with the final exfoliation step performed using cross-linked adhesive polymer film (Gel-pak® 4x adhesion) that does not leave surface residues. Single-, double- and few-layer MoS$_2$ flakes were transferred onto uncoated and Au-coated (5 nm Cr/40 nm Au) Si/SiO$_2$ substrates with 300 nm thick thermal silicon oxide on doped Si substrate. The substrates had narrow (150 to 200 nm wide) trenches through all depth of SiO$_2$ produced using optical edge lithography described elsewhere [34]. Prior to use, substrates were cleaned in piranha solution (3:1 concentrated H$_2$SO$_4$ to 30% H$_2$O$_2$) and immediately before transfer by 98% Ar/2% oxygen plasma (PlasmaPrep2, Gala Instruments) for three minutes to remove any remaining organic material and facilitate attachment of the MoS$_2$ to the surface.

**Scanning Kelvin probe microscopy (SKPM)**

The surface potential ($V_{CPD}$) measurements have been performed by SKPM, which also provided information on sample morphology, as well as a quantitative determination of the local thickness of MoS$_2$. The thickness of the flake was defined using the histogram method, allowing the highest accuracy of measurements of thin layers [35]. SKPM measurements were conducted in ambient conditions, on a Bruker Icon AFM, using Bruker highly doped Si probes (PFQNE-AL) with a force constant ~ 0.9 N/m. Frequency-modulated SKPM (FM-SKPM) technique operated in a single pass mode has been used in all measurements. FM-SKPM operates by detecting the force gradient ($dF/dz$), which results in changes to the resonance frequency of the cantilever. In this technique, an AC voltage with a lower frequency ($f_{mod}$= 3 kHz) than that of the resonant frequency ($f_0$= 300 kHz) of the cantilever is applied to the probe, inducing a frequency shift. The feedback loop of FM-KPFM monitors the side bands, $f_0 \pm f_{mod}$, of cantilever vibration and minimises them by applying an offset DC voltage which is recorded to obtain a surface potential map [36]. Sample-biased setup has been used in the present work.

**Raman spectroscopy**

Raman intensity maps were obtained using a Horiba Jobin-Yvon HR800 System. A 532-nm wavelength laser (2.33 eV excitation energy) was focused onto the sample through a 100 × objective with Raman maps data taken with a spectral resolution of (3.1 ± 0.4) cm$^{-1}$ and lateral spatial resolution of (0.4 ± 0.1) μm. The lateral spatial resolution was determined using Si calibration gratings with a period of 3 μm from the first derivative of Si 520 cm$^{-1}$ band scan lines extracted from the Raman map. The width of the 1st derivative peak obtained by fitting with a Gaussian is the lateral resolution.

Laser power was kept to a minimum to avoid heat-induced damage to the sample [37, 38], which can also lead to broadening and shifting of the Raman peaks. Raman intensity maps for E$^1_{2g}$ and A$_{1g}$ modes were collected simultaneously by taking individual spectra on preselected areas across each flake. The Raman spectra were processed afterwards using Matlab software to fit the E$^1_{2g}$ and A$_{1g}$ peaks using single Lorentzians and the resulting frequency shifts plotted to yield frequency shift maps.

**Ultrasonic force microscopy (UFM)**

UFM has previously shown to be highly sensitive to both the surface, subsurface and layer-interface structure of $MoS_2$ [4], graphene [32] and other 2D materials [39]. UFM is a modification of standard contact mode AFM where the sample is oscillated at low amplitude (5-10 Å) and high frequency (4-5 MHz in our measurements). At these frequencies the cantilever (Contact-G, Budget Sensors) probes becomes dynamically extremely rigid resulting in periodic indentation of and separation from the sample. Resulting detection of the ultrasonic vibration via non-linear interaction forces between the probe and sample during an oscillation cycle is then strongly dependent upon the local mechanical structure of the sample [40] allowing effective non-destructive mapping of the nanomechanical properties with a lateral resolution of 2-3 nm.

**Results and Discussion**

**Functional scanning probe microscopy**

Topography image of the exfoliated and transferred $MoS_2$ flake on a gold substrate, containing regions of different thicknesses, is shown in figure 1(a). Thickness of individual regions has been defined using tapping mode AFM, where the 1$^{st}$ layer on the substrate has a thickness of ~1 nm. Consequent layers thicknesses have been estimated from a topographic line profile (figure 1(d)) considering an interlayer separation of 0.7 nm, yielding, correspondingly, 1, 5, 8 layers thickness and bulk (~40 layers). The vertical trench in the substrate is highlighted by an arrow.

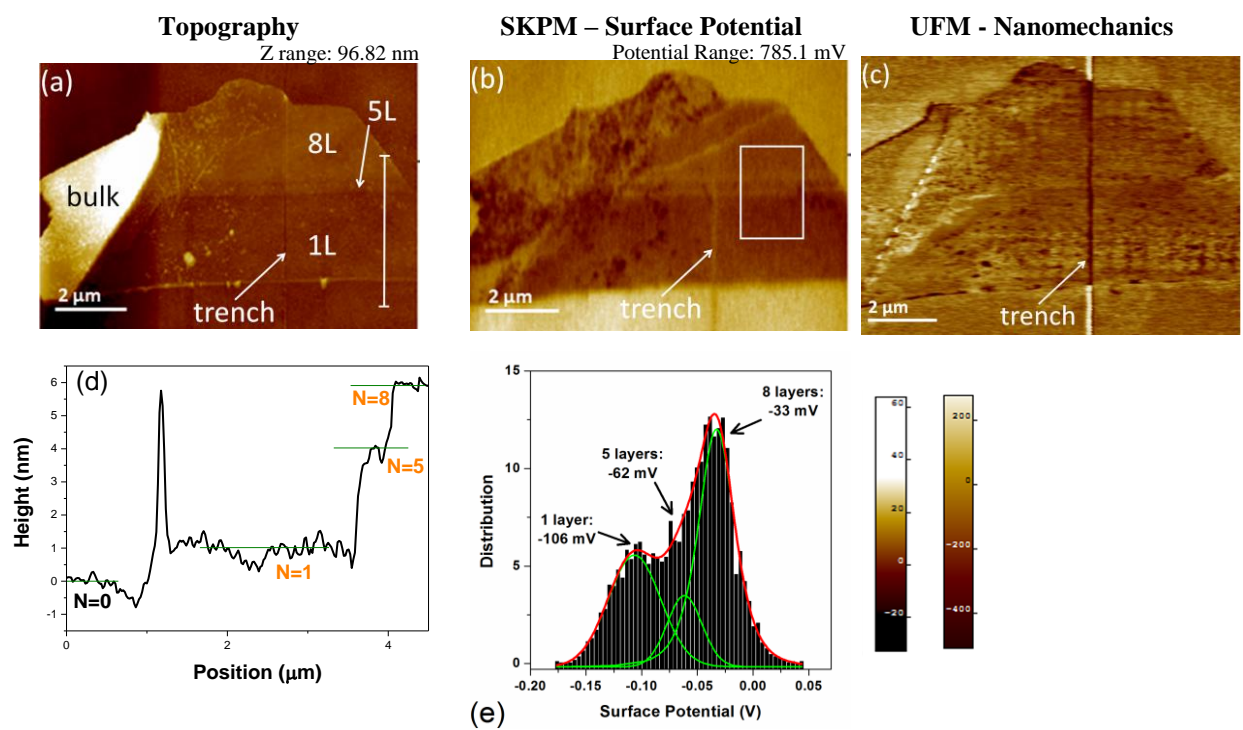

**Figure 1.** (a) Topography, (b) SKPM surface potential and (c) UFM nanomechanical images of the $MoS_2$ flake on Au substrate, the number of layers is indicated in (a). (d) Topography profile along the line indicated in (a). (e) Distribution of surface potential across regions of different thickness of $MoS_2$ as indicated by the frame in (b). The green lines are result of the fitting for individual domains, the red line shows the overall fit.

Surface potential mapping was obtained simultaneously with topography imaging and is presented in figure 1(b). For the entire MoS$_2$ flake, $V_{CPD}$ value is significantly lower than that of the gold substrate, where $V_{CPD}$ value of a single layer is notably the lowest. The $V_{CPD}$ value increases with the layer thickness, consistent with literature [24], though the bulk value (as well as a part of the flake) is compromised by decoration of the surface by environmental adsorbates (also seen as bright clusters in figure 1(a)), leading to a decrease in surface potential of the MoS$_2$ layers [24]. With this exception, distribution of the surface potential within each layer is relatively homogeneous within the 20 mV accuracy. Figure 1(e) demonstrates the distribution of the surface potential across an area of different thicknesses indicated by the frame in figure 1(b) and a histogram analysis of the acquired SKPM data over the enclosed area is presented next. The average values of the measured surface potentials on the different thickness domains are obtained by peak fitting, as indicated by the green lines in Figure 1(e), showing the individual resulting components. The results are best described by the fitting of three Lorentzian peaks, corresponding to the areas of 1, 5 and 8 MoS$_2$ layers with their absolute $V_{CPD}$ values being -106, -62 and -33 mV, respectively. The assignment of individual domains thicknesses to the deconvoluted peaks in the histogram has been made in correlation with the level of contrast in the associated SKPM image, with the most intense peak corresponding to the brightest contrast in the surface potential image.

Using these absolute values for the contact potential difference and a work function of 4.5 eV for the scanning tip [41], the work function can be estimated according to $\Phi_{sample} = \Phi_{tip} + V_{CPD}$, resulting in 4.39 eV, 4.44 eV and 4.47 eV for 1, 5, and 8 layers, respectively, consistent with the values reported by Ochedowski et al. [24]. However, a significantly larger value (up to 5.25 eV) for bulk MoS$_2$ was used by others, see *e.g.* Ref. [23]. It should also be noted that although an opposite thickness-dependent trend of the surface potential was experimentally observed in Ref. [22, 23], this discrepancy arises from the use of different type of biasing (tip or sample biasing) in the SKPM setup.

UFM-derived nanomechanical mapping (figure 1(c)) shows stiffness variations arising from both the MoS$_2$ flake thickness, the local sample-substrate interface and the interlayer coupling; these sources may be interlinked, for example significantly more variation are observed in the monolayer region compared to the thicker areas of material (figure 1(c)), where brighter contrast corresponds to more mechanically stiff areas. Here the trench can be clearly seen as the black line (zero signal) running across the flake under its surface; zero UFM signal indicates complete local decoupling of the sample from the substrate and allows us to precisely mark the transition from supported to suspended MoS$_2$. Small dark points on the monolayer region correspond to surface contamination of the flake and not delaminated MoS$_2$ (bright trench contrast outside of flake MoS$_2$ footprint is an artefact associated with a steep and high edge). Regions of different thickness are easily identified and generally uniform UFM response within each region suggests an excellent MoS$_2$-substrate contact.

Observed in all imaging modes (figure 1 (a-c)) the contamination at the bulk part of the flake arises from absorbed water and other airborne molecules, as MoS$_2$ is mildly hydrophilic and sensitive to atmospheric water vapour. Annealing of the flake led only to partial disappearance and redistribution of this contamination. Due to MoS$_2$ polarity and hydrophilicity, the effect of atmospheric species can be rather strong, *e.g.* through the formation of hydrogen bond with water molecules. Such adsorbates act as charge trappers, affecting the surface potential and charge distribution at ambient atmosphere. A cleaning procedure using contact AFM and soft cantilevers, similar to the one described in Goosens et al. [43], has been employed, however no significant improvement in the flake appearance was noticed after a few cleaning cycles.

**Raman mapping: thickness dependence**

Raman maps of $E^1_{2g}$ and $A_{1g}$ modes were obtained using 532-nm laser line for a flake containing domains of 1L, 2L, 3L and 7L thickness on Si (figure 2). Corresponding maps of the frequency shift and intensity are shown in figures 2 (a,b) and 2 (c,d), respectively. The maps demonstrate that both the frequency shift and intensity maps are thickness-dependent, allowing one to use such maps for a clear identification of the layer number.

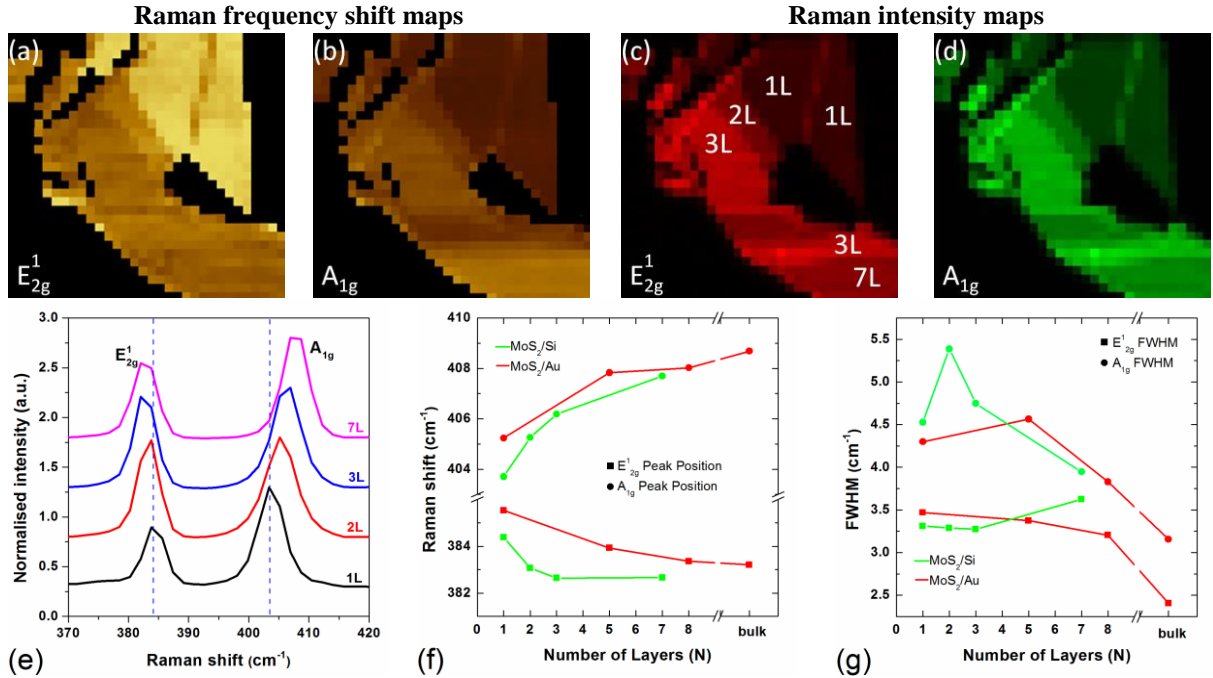

**Figure 2.** Raman frequency shift (a, b) and intensity (c, d) maps of $E^1_{2g}$ (a, c) and $A_{1g}$ (b, d) peaks for the MoS$_2$/Si flake with 1, 2, 3, and 7 layers domains. (e) Representative Raman spectra for different thicknesses of the MoS$_2$ flake. (f) Thickness-dependent Raman shifts of $E^1_{2g}$ and $A_{1g}$ modes for MoS$_2$ flakes on Au and Si substrates. (g) Line widths of $E^1_{2g}$ and $A_{1g}$ modes for MoS$_2$ on Au and Si substrates as a function of the number of layers.

Spatial maps of the Raman frequency shift for the $E^1_{2g}$ and $A_{1g}$ modes (figures 2 (a, b)) demonstrate that the $E^1_{2g}$ vibration softens (red shift), while the $A_{1g}$ vibration stiffens (blue shift) with increasing sample thickness. For example, in figure 2(a) the $E^1_{2g}$ mode has lowest contrast for 1L thickness and highest for the thicker material, indicating the red shift for the latter. Conversely, the opposite trend is observed for the $A_{1g}$ mode map (figure 2 (b)), where the 1L domain has the brightest and the bulk area – the darkest contrast, indicative of the blue shift for the thicker material.

In order to further quantify Raman maps, we extract individual Raman spectra obtained on flakes of different thickness. Typical Raman spectra highlighting the region of $E^1_{2g}$ and $A_{1g}$ modes are shown in figure 2 (e). The dependence of peak position on the number of layers is plotted in figure 2 (f). The Raman shift between $E^1_{2g}$ and $A_{1g}$ modes, which becomes generally larger with layer number, is in good agreement with literature [13-17]. Intensity maps of $E^1_{2g}$ and $A_{1g}$ modes are shown in figures 2 (c) and 2 (d), respectively. For both types of vibrations, the intensity increases with thickness between 1L through to 7L. The widths of the peaks are generally independent on the layer thickness for small number of layers, with peaks becoming narrower for bulk MoS$_2$, also in agreement with previous results [14].

Comparison of the influence of a metallic substrate (Au), relative to the semiconducting one, on the vibrational properties of the MoS$_2$ is summarised in figure 2 (f). Here, the difference between the E$^1_{2g}$ and A$_{1g}$ peak positions for 1L is comparable for both substrates: $\Delta\omega_{Au}$ = 19.7 cm$^{-1}$ and $\Delta\omega_{Si}$ = 19.3 cm$^{-1}$. However, a noticeable red shift is observed for both E$^1_{2g}$ and A$_{1g}$ modes when MoS$_2$ is deposited on Si in comparison with the Au substrate. This observation is consistent with a more strained MoS$_2$ on the Si substrate compared to the Au one, as it was shown that tensile strain is associated with a red shift of the Raman E$^1_{2g}$ mode for monolayer MoS$_2$ [24]. At the same time, for single layer MoS$_2$ on Au, the A$_{1g}$ peak becomes sharper and E$^1_{2g}$ broader than for single layer MoS$_2$ on Si. It would be interesting to examine how the surface potential is affected by strain of the flake, however due to the semi-insulating nature of the Si substrate giving rise to charging of the flake during scanning, SKPM data has only been acquired on the Au substrate, so a direct comparison is not possible based on the current data.

**Raman mapping: comparison of supported vs suspended MoS$_2$**

Furthermore, we studied the effects on the vibrational spectra and surface potential of the different MoS$_2$ thickness regions arising from the supported (on Au substrate) or suspended (over a channel in the same substrate) nature of the flake. For this, we compare the Raman spectra of supported and suspended 1L and 8L MoS$_2$ regions (figure 3) of the same flake as shown in figure 1. Corresponding images for Raman intensity and shift are shown in figure 3 (a, b) and 3 (c, d), respectively. Overall, the image demonstrates the same thickness-dependent tendencies as discussed above for MoS$_2$ on Si. However, the trend is not observed for the E$^1_{2g}$ mode in the bulk domain (figure 3 (c)), which might be explained by contamination of the representative area. In order to elucidate the effect of the substrate, we compare averaged Raman spectra taken over identical areas (highlighted by the frames in figure 3 (d)) on 1L and 8L flakes and positioned either on the supported part of the flake or over the channel (see schematics in figure 3 (d)). For consistency, the area was kept constant for all considered cases. Characteristic Raman spectra are presented in figure 3 (e). An increase in intensity is observed for suspended regions with respect to the supported ones, as seen for both E$^1_{2g}$, A$_{1g}$ modes, but more pronounced for the out-of-plane one. It should be pointed out that the channel in the substrate is narrower than the lateral spatial resolution of the Raman system and so, the resultant spectra show a mixed contribution from the material suspended over the channel and adjacent regions supported on the Au substrate. Raman intensities are known to be sensitive to the orientation of single crystals in relation to the scattering geometry [42]. Since our measurements are averaged over an area larger than the width of the channel, a misalignment of the crystal plane for the suspended MoS$_2$ in respect to the area of the flake supported by the Au substrate is unlikely to be solely accounted for the increase in intensity we observe. Another possible reason could be due to interference effects caused by multiple reflections of the laser beam at the edges of the channel, as it is well recognized that optical interference effects have a strong impact on the intensity of Raman spectra, as shown e.g. for graphene [46]. Contrary to our observations, for monolayer MoS$_2$, calculations of the enhancement factor of Raman peak intensity due to optical interference effects show a stronger Raman response for supported configurations than for suspended ones [47]. This could point to a different mechanism giving rise to the increase in intensity we observe, possibly based on emission and absorption effects at the edges of the channel and the metallic nature of the substrate as compared to the dielectric ones usually employed for graphene and other 2D materials. Furthermore, we also observe an increased Raman signal at the channel edge in the bare substrate not associated with the flake, which could be an indication of SERS-like enhancement, as many nanoscale-textured metallic surfaces have been found capable of producing strong electromagnetic fields that give rise to SERS enhancements.

The enhancement could be caused by the polarization of the incident laser light possibly resonating with the narrow Au channel to give an enhancement of the absorption and scattering processes cross sections [48].

However, no notable change in the frequency shift has been measured over the suspended areas relative to the supported ones.

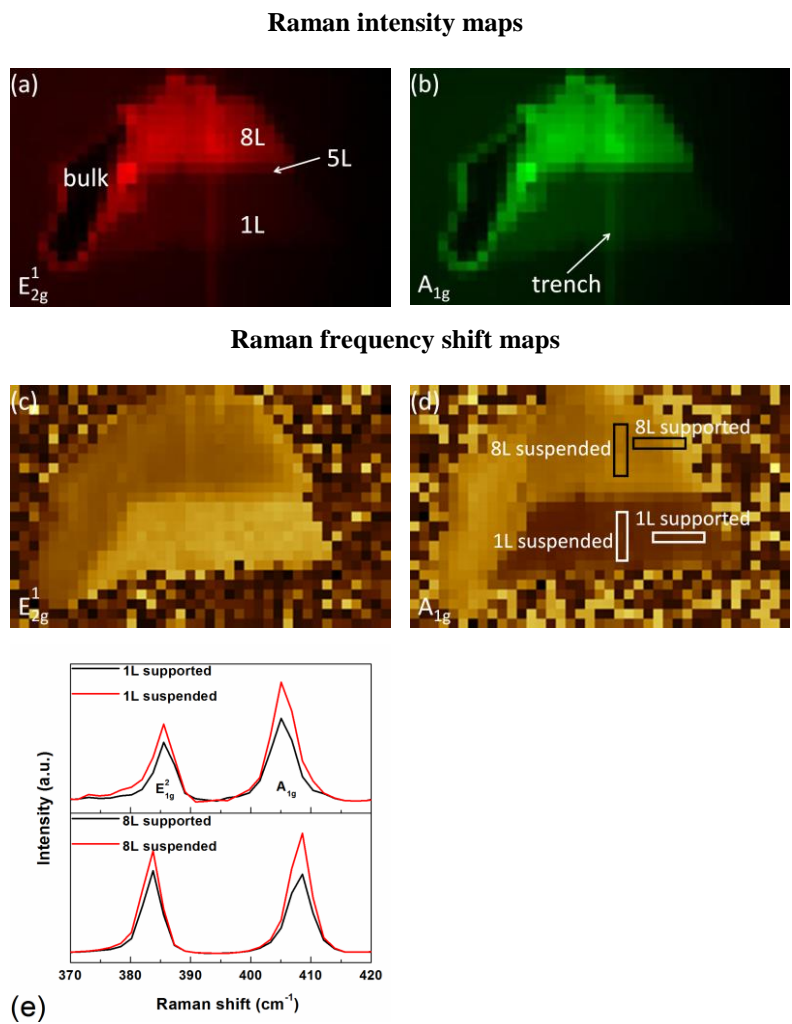

**Figure 3.** Raman intensity (a, b) and frequency shift (c, d) maps of $E^1_{2g}$ (a, c) and $A_{1g}$ (b, d) peaks for the MoS$_2$/Au flake with 1, 5, 8 layers and bulk domains. (e) Comparison of Raman spectra for 1L and 8L domains taken from the areas highlighted in (d).

Raman maps also demonstrate that the signal for both vibrational modes, while being generally homogeneous within each individual supported flake, shows a change of the contrast at the flake border. This is manifested as an increase of intensity of both $E^1_{2g}$ and $A_{1g}$ modes. The effect is evident for thicker flakes, however is less pronounced for 1L MoS$_2$. For Raman shifts the behaviour is more complex, *i.e.* the $E^1_{2g}$ mode is characterised by a darker contrast (*i.e.* experiences a blue shift), whereas the $A_{1g}$ mode demonstrates a brighter one (*i.e.* experiences a red shift) at the edges of individual flakes. This behaviour is opposite to the thickness-dependent trends as observed above and likely to reflect the defective nature and possible inhomogeneity of the chemical composition of the flake boundaries due to adsorption of adatoms and creation of vacancy defects [43, 44].

Comparison of UFM and SKPM responses for the different MoS$_2$ layer thicknesses (1L, 5L and 8L areas) over the supported/suspended transition region (figure 4 (a-c)), shows a clear peak (enhanced surface potential contrast) in the surface potential (blue dashed lines) over the UFM identified trench (black solid line) for the 1L regions, smaller changes are observed for the thicker 5L and 8L regions. For comparison, figure 4 (d) demonstrates the increase of the Raman intensity for both main modes over the same region of MoS$_2$.

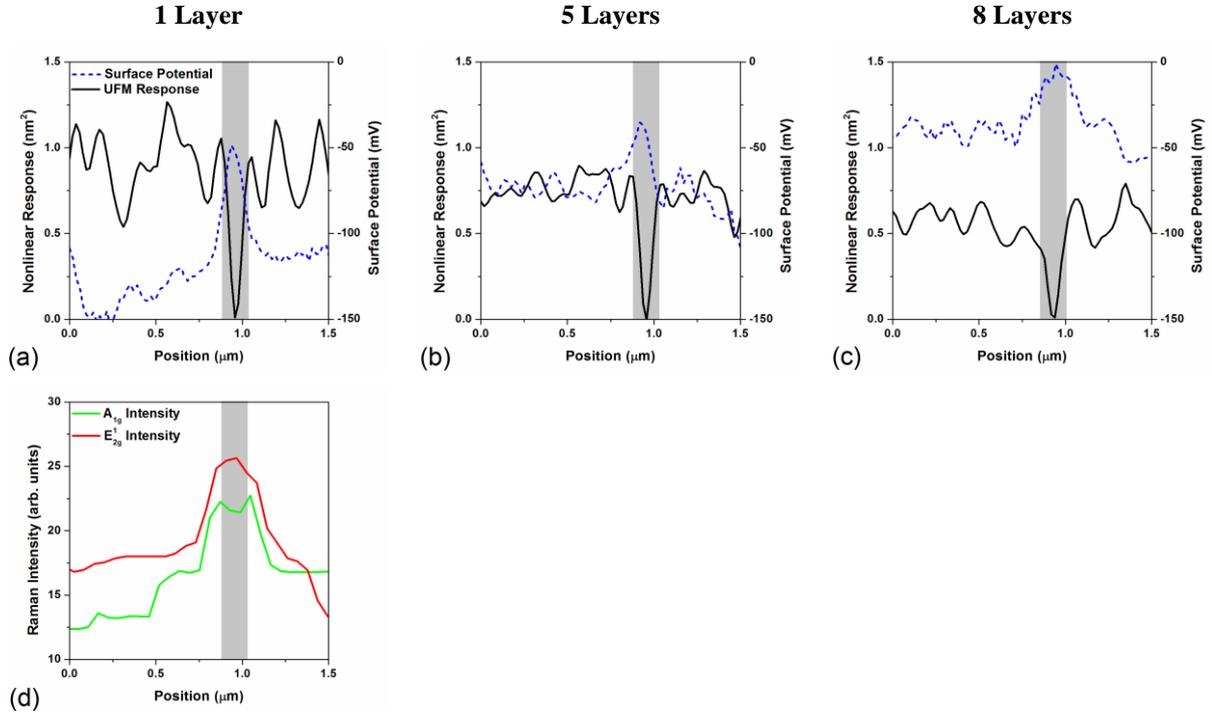

**Figure 4.** Comparison of UFM nanomechanical (black solid lines) and SKPM surface potential (blue dashed lines) responses in the 1 layer (a); 5 layer (b) and 8 layer (c) regions and corresponding traces across the Raman intensity maps (d) for the $E^1_{2g}$ and $A_{1g}$ modes (red and green lines, corresponding to maps in figures 3(a) and (b) respectively). The position of the trench is represented by the grey shaded area.

Surface potential data arising from comparison of the supported and suspended regions are shown in Table I. It is noteworthy that the area of unsupported single layer MoS$_2$ above the channel is characterised by an enhanced surface potential contrast, *i.e.* ~ 100 mV increase as compared to the main part of a single flake. This is most likely due to suppressed charge transfer for the suspended monolayer compared to the supported one [22, 24]. The effect of the trench is significantly less pronounced in the case of thicker (5L and 8L) flakes.

**Table I.** Average enhanced surface potential contrasts between the supported and suspended regions of the MoS$_2$.

| MoS$_2$ thickness | Increase in surface potential contrast | |
|---|---|---|
| | Absolute change | % change |
| 1 Layer | 94 ± 7 mV | (67 ± 5%) |
| 5 Layers | 39 ± 17 mV | (53 ± 23%) |
| 8 Layers | 41 ± 4 mV | (98 ± 10%) |

The width of the trench as observed by UFM is typically ~150 nm for both the 1 and 5 layer regions and ~190 for the 8 layer region, in good correlation with the SKPM surface potential peaks. Apparent broadening of the trench in thicker regions is well known due to increased mechanical rigidity of the thicker flake. However, for $E^1_{2g}$ and $A_{1g}$ Raman modes the increased intensity corresponding to the suspended region is in the range 410-540 nm. The broader response profile in figure 4(d) is due to lower spatial resolution of the Raman spectroscopy as compared with AFM-based techniques. As a result, mixing of the signal from suspended and supported areas occurs and the possible effect of the channel might be reduced.

**Conclusions**

In conclusion, we have performed a comprehensive study of the optical, nanomechanical and electrostatic properties of $MoS_2$ in dependence on the layer thickness and sample-substrate interaction. Using subsurface sensitive UFM mapping we have identified the change in the nanomechanical properties of the $MoS_2$ flake indicating effect of the substrate for a range of flake thicknesses (1, 5 and 8 layers) and correlated these regions with SKPM derived surface potential and Raman mapping. We observe an increase in the surface potential contrast for suspended regions of all thicknesses relative to the supported areas, with the monolayer region demonstrating a ~100 mV (~67%) increase, which is believed to be due to suppressed charge transfer for the suspended monolayer compared to the supported one. Furthermore, a corresponding increase in Raman intensity for the $E^1_{2g}$ and $A_{1g}$ modes is observed for the monolayer region but not for thicker regions of the flake. This thickness-dependent enhancement arises due to a different crystal orientation on the suspended area compared to the supported region of the monolayer flake. Additionally, we demonstrate a noticeable red shift for both $E^1_{2g}$ and $A_{1g}$ modes when $MoS_2$ is deposited on Si in comparison with the Au substrate. This observation is consistent with a more strained state of the $MoS_2$ flake on the Si substrate.

These results provide a detailed understanding of the layer properties, which are essential for potential optoelectronic applications by decoupling the optical and electrostatic properties of $MoS_2$ from substrate-induced effects.

**Acknowledgements**

Authors acknowledge support of EC grants Graphene Flagship (No. CNECT-ICT-604391), FUNPROB, GRENADA, and EMRP under project GraphOhm, EPSRC grants EP/G015570 and EP/K023373/1, NMS under the IRD Graphene Project and NowNANO Doctoral Training Centre. Authors thank Dr Spyros Yannopoulos for useful discussions on Raman spectroscopy and Dr Mark Rosamond and Dr Dagou Zeze for supplying the trenched substrates.